\documentclass{ws-procs961x669}

\newcommand{\old}[1]{}

\begin{document}

\title{Magnetized tori around a uniformly accelerating black hole}

\author{Shokoufe Faraji, Audrey Trova}

\address{%
 University of Bremen, Center of Applied Space Technology and Microgravity (ZARM), 28359 Germany
}%

\begin{abstract}
We generalize the relativistic accretion thick disc model to the background of a spinning charged accelerating black hole described by the C-metric to study the effects of this background on the disc model. We show the properties of this accretion disc model and its dependence on the initial parameters. This background can be distinguishable from the Kerr space-time by analyzing the observing features of accretion discs.
\end{abstract}


\bodymatter

\section{Introduction}
In this paper we investigate the spinning charged C-metric, with the aim of studying the properties of the well known magnetised Thick accretion disc model and the morphology of the equipotential surfaces. 

The family of C-metric has accelerating nature and is considered as describing an accelerating black hole \citep{PhysRevD.2.1359}. The spinning charged C-metric in Boyer-Lindquist-type coordinates \citep{2005CQGra..22..109H} reads as  

\begin{align}\label{eq:metric}
&ds^2=\frac{1}{\Omega^2}\left(-\frac{f}{\Sigma}\left[dt-a\sin^2\theta\frac{d\varphi}{K}\right]^2\right.\nonumber\\
&\left.+\frac{\Sigma}{f}dr^2+\Sigma r^2\frac{d\theta^2}{g}+\frac{g\sin^2\theta}{\Sigma r^2}\left[adt-(r^2+a^2)\frac{d\varphi}{K}\right]^2\right),\  
\end{align} 
where

\begin{align}
    &\Omega=1+\alpha r \cos\theta,\\
    &f(r)=\left(1-\alpha^2r^2\right)\left(1-\frac{2m}{r}+\frac{e^2+a^2}{r^2}\right),\\
    &g(\theta)=(e^2+a^2)\alpha^2 \cos^2\theta+2m\alpha \cos\theta+1,\\
    &\Sigma(r,\theta) = \frac{a^2}{r^2}\cos^2\theta+1,\\
    &\xi=\alpha^2(e^2+a^2)+1,\\
    &K=\xi+2m\alpha,
\end{align}
where $t\in (-\infty,+\infty)$, $\theta\in (0, \pi)$, $r\in (0, +\infty)$. The metric has four independent parameters: the mass $m$, the electric charge $e$, the rotation $a$, and the so-called acceleration parameter $\alpha$. The spacetime has a conical
singularity along one, or both polar axes; conical surplus angle at $\theta=0$ and a deficit angle at $\theta=\pi$. The parameter $K$ regularizes the distribution of conical defects in the space-time and allows $\varphi$ to be $2\pi$-periodic (see for example \cite{doi:10.1063/1.523896,1999PhRvD..60d4004B,2006CQGra..23.6745G}).

Almost all studies with this metric are revolved around the coordinate ranges, which are dictated by the metric functions and their root configurations. The main constrains are 

\begin{align}
  &r_b = -\frac{1}{\alpha\cos \theta},\\
   &e^2+a^2\leq m^2,  \
\end{align}
 and 
 \begin{align}\label{regioneq}
 2m\alpha\leq
   \left\{
  \begin{array}{@{}ll@{}}
  2\sqrt{\xi-1} & \xi>2, \\
    \xi & 0<\xi\leq 2.
    \end{array}\right.
\end{align}

\section{Properties of the disc model }

The Thick accretion disc is an analytical model that provides a general method to build equilibrium configurations of the perfect fluid matter orbiting around a stationary and axially symmetric black hole \citep{1989rfmw.book.....A,1978srfm.book.....D}. In this procedure, we considered the approach of Komissarov \citep{Komissarov_2006} to attach a dynamically toroidal magnetic field to the model (for more details see \cite{2022PhRvD.105j3017F}).

An analysis shows \cite{2021arXiv210807070F} that by increasing the acceleration parameter $\alpha$, almost the possibility of having solutions decreases dramatically. On the contrary, the charge parameter $e$ positively contributes to having solutions; mainly, its effect manifests more for a relatively large $\alpha$. Furthermore, the possibility of having solutions for larger $\alpha$ depends on having large values for $e$, so the effect of higher $\alpha$ on the disc could be neutralized partially with the higher charge.

\begin{figure*}
  \centering
\begin{tabular}{ccc}
\includegraphics[width=4.2cm]{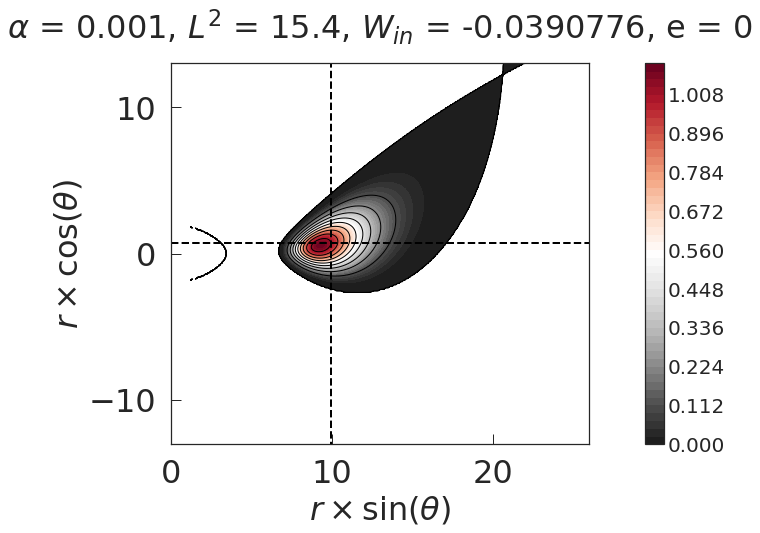}&
\includegraphics[width=4.2cm]{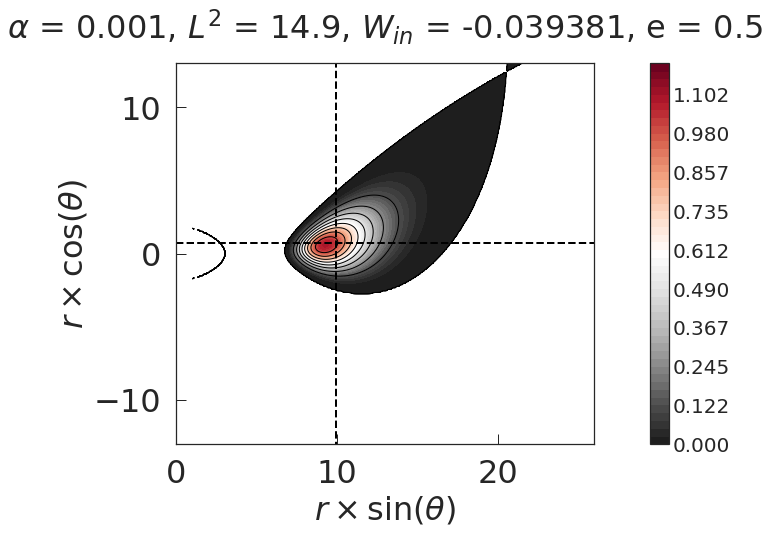}&
\includegraphics[width=4.4cm]{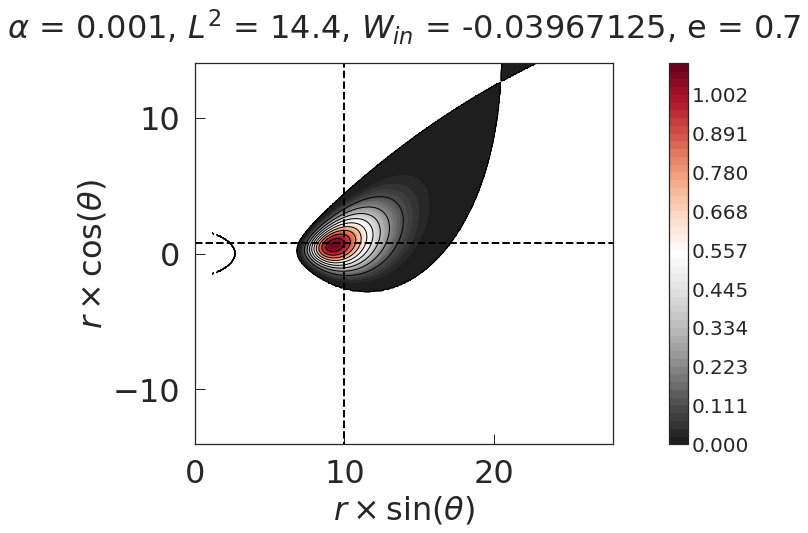}
\end{tabular}
    \caption{Contour map of the rest-mass density of magnetised disc. The dashed lines point the center of the disc at $r_c=10$. The discs are all highly magnetised disc.}
   \label{fig:figsolre2} 
\end{figure*}

The effect of rotation parameter $a$ on having solutions is not strong compared to $\alpha$ but more substantial than the charge parameter $e$. In fact, parameter $a$, like $\alpha$, has a negative effect on having solutions. In general, as $e$ increases, we expect the matter is concentrated closer to the inner edge of the disc. On the contrary, the higher values of $a$ spread the matter more through the disc. Moreover, Increasing $a$ decreases the size of the disc and change the distribution of matter. In addition, it shifts the disc farther from the compact object, contrary to an increase in $e$, which shift the disc closer to the central object. 


For a fixed value of acceleration parameter $\alpha$ and vanishing rotation, the magnetization parameter does not influence the geometry of the disc; however, it changes the distribution of matter inside the disc and shifts the location of the rest-mass density maximum. In addition, we have a larger oriented disc for larger values of $e$.

\begin{figure*}
  \centering
\begin{tabular}{cc}
\includegraphics[width=6cm]{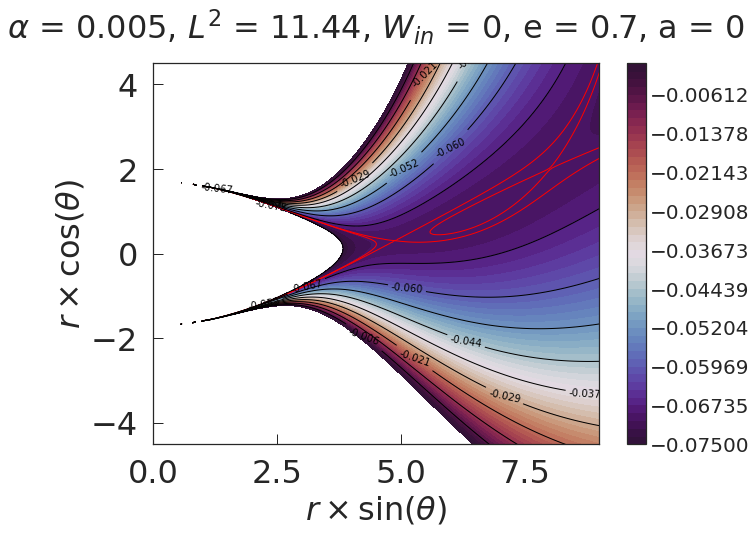}&
\includegraphics[width=6.3cm]{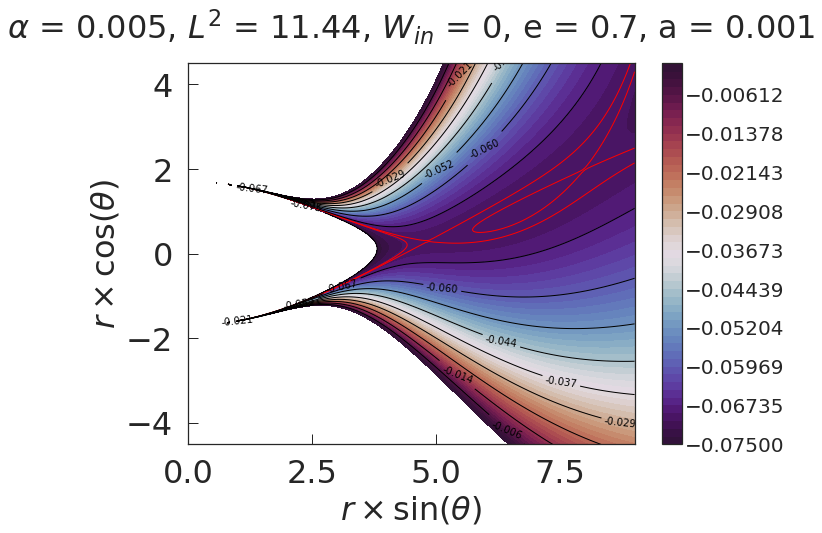}\\
\includegraphics[width=6cm]{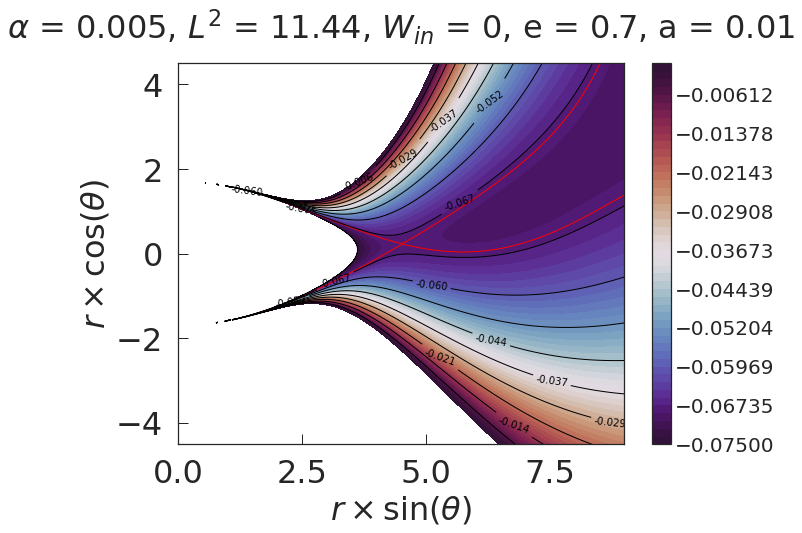}&
\includegraphics[width=6cm]{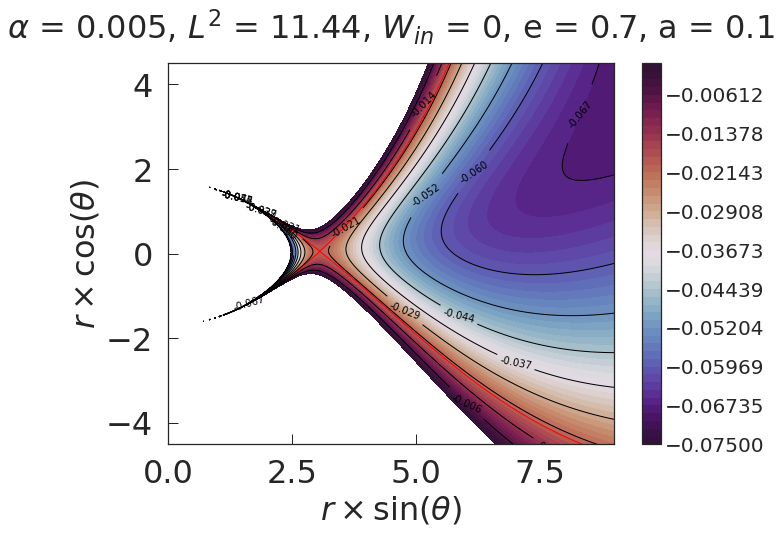}
\end{tabular}
    \caption{Contour map of the equipotential surfaces. no solution for $\alpha=0.005$ unless for high $e$ and very low $a$. Show no solution for higher $\alpha$ and rotation.}
    \label{fig:figsolre3rot} 
\end{figure*}


\section{Conclusion}\label{sec5}

In this work, we discussed the influence of the magnetisation parameter, charge $e$, rotation $a$, and accelerating parameter $\alpha$ in the spinning charged C-metric on the structure of the magnetised thick disc model. 

In general, we can have the Thick disc model for relatively small values of $\alpha$, while by increasing $\alpha$, the disc structure becomes smaller and gradually vanishes. The same is true for $a$. On the contrary to these two parameters, an increase in $e$ increases the disc size and possibility of having a solution. However, we should mention that the strength of the parameters are not the same, as $\alpha$ has the strongest and $e$ has the weaker effect on the disc structure, in comparison.

A further step of this work can be a study on the oscillation of the disc in this setup which is in progress. In addition, this is interesting to examine the Thick disc model with different angular momentum profiles in this space-time.

\bibliographystyle{ws-procs961x669}
\bibliography{discCmetric.bib}

\begin{thebibliography}{10}

\bibitem{PhysRevD.2.1359}
W.~Kinnersley and M.~Walker, Uniformly accelerating charged mass in general
  relativity, {\em Phys. Rev. D} {\bf 2}, 1359 (Oct 1970).

\bibitem{2005CQGra..22..109H}
K.~{Hong} and E.~{Teo}, {A new form of the rotating C-metric}, {\em Classical
  and Quantum Gravity} {\bf 22}, 109 (January 2005).

\bibitem{doi:10.1063/1.523896}
F.~J. Ernst, Generalized c‐metric, {\em Journal of Mathematical Physics} {\bf
  19}, 1986  (1978).

\bibitem{1999PhRvD..60d4004B}
J.~{Bi{\v{c}}{\'a}k} and V.~{Pravda}, {Spinning C metric: Radiative spacetime
  with accelerating, rotating black holes}, {\em \prd} {\bf 60}, p. 044004
  (August 1999).

\bibitem{2006CQGra..23.6745G}
J.~B. {Griffiths}, P.~{Krtous} and J.~{Podolsk{\'y}}, {Interpreting the
  C-metric}, {\em Classical and Quantum Gravity} {\bf 23}, 6745 (December
  2006).

\bibitem{1989rfmw.book.....A}
A.~M. {Anile}, {\em {Relativistic fluids and magneto-fluids : with applications
  in astrophysics and plasma physics}} (Oxford Univ. Press, 1989).

\bibitem{1978srfm.book.....D}
W.~G. {Dixon}, {\em {Special relativity: the foundation of macroscopic
  physics.}} (Oxford Univ. Press, 1978).

\bibitem{Komissarov_2006}
S.~S. Komissarov, Magnetized tori around kerr black holes: analytic solutions
  with a toroidal magnetic field, {\em Monthly Notices of the Royal
  Astronomical Society} {\bf 368}, p. 993–1000 (Apr 2006).

\bibitem{2022PhRvD.105j3017F}
S.~{Faraji}, A.~{Trova} and V.~{Karas}, {Magnetized relativistic accretion disk
  around a spinning, electrically charged, accelerating black hole: Case of the
  C metric}, {\em \prd} {\bf 105}, p. 103017 (May 2022).

\bibitem{2021arXiv210807070F}
S.~{Faraji} and A.~{Trova}, {Magnetised relativistic accretion disc around a
  spinning charged accelerating black hole}, {\em arXiv e-prints} , p.
  arXiv:2108.07070 (August 2021).

\end{thebibliography}

\end{document}